# Facile and Fast Transformation of Non-Luminescent to Highly Luminescent MOFs: Acetone Sensing for Diabetes Diagnosis and Lead Capture from Polluted Water

*Mario Gutiérrez,[a] Annika F. Möslein,[a] and Jin-Chong Tan [a]\**

[a] Multifunctional Materials & Composites (MMC) Laboratory, Department of Engineering Science, University of Oxford, Parks Road, Oxford OX1 3PJ, United Kingdom.

**Abstract**

Metal-organic frameworks (MOFs) stand as one of the most promising materials for the development of advanced technologies owing to their unique combination of properties. The conventional synthesis of MOFs involves a direct reaction of the organic linkers and metal salts, however, their post-synthetic modification is a sophisticated route to produce new materials or to confer novel properties that cannot be attained through traditional methods. This work describes the post-synthetic MOF-to-MOF transformation of a non-luminescent MOF (Zn-based OX-1) into a highly luminescent framework (Ag-based OX-2) by a simple immersion of the former in a silver salt solution. The conversion mechanism exploits the uncoordinated oxygen atoms of terephthalate linkers found in OX-1, instead of the unsaturated metal sites commonly employed, making the reaction much faster. The materials derived from the OX-1 to OX-2 transformation are highly luminescent and exhibit a selective response to acetone, turning them into a promising candidate for manufacturing fluorometric sensors for the diagnosis and monitoring of diabetes mellitus. Our methodology can be extended to other metals such as lead (Pb). The fabrication of a polymer mixed-matrix membrane containing OX-1 is used as a proof-of-concept for capturing Pb ions (as pollutants) from water. This research instigates the exploration of alternative methodologies to confer MOFs with special aptitudes for photochemical sensing or for environmental applications such as water purification.

**Keywords:** MOF-to-MOF transformation, Luminescence, Acetone Sensing, Diabetes Monitoring, Pb capture, Water Purification.

**Introduction**

Metal-organic frameworks (MOFs) are a class of porous ordered structures constructed from a periodic self-assembly of organic and inorganic subunits.[1] Over the last two decades, MOFs have received great attention from a broad interdisciplinary number of scientists, including chemists, physicists, material scientists or engineers among others. The versatile physicochemical properties of MOFs make them potential candidates for emerging key technologies.[2-4]

MOFs are typically synthesized by a direct reaction of the organic linkers and the metal salts through numerous synthetic approaches.[5] However, during the past few years, distinct routes have been investigated in order to develop new MOF materials with striking properties that could not be achieved through conventional methods. One of the most deeply explored mechanism is the post-synthetic modification of MOFs.[6-8] Although most of the examples were based on the modification of the organic linkers,[9-11] an increasing number of cases employing metal exchange have been reported.[12, 13] The post-synthetic modifications by leveraging the metal exchange approach can be accomplished through a charge balanced substitution of the cations present in the pores, or by a more sophisticated process which encompasses the rupture and formation of coordination bonds.[14] Therefore, the cation exchange process will be more effective in those metals containing coordinatively unsaturated sites, or metals with saturated positions that can still undergo cation exchange as a result of weak field interactions with the organic linkers.[13] The coexistence of multiple metals in a single MOF crystal generates a subclass of materials known as the mixed-metal MOFs (MM-MOFs).[15] Some MM-MOFs have exhibited improved performance in catalysis,[16, 17] photocatalysis[18, 19] and gas adsorption[20, 21] technologies, while others have also shown outstanding luminescent properties, making them promising candidates to be deployed in the fabrication of disruptive devices, such as luminescent thermometers,[22, 23] and optical sensors.[24, 25]

The substitution of metal cations in MOFs can also induce a complete MOF-to-MOF transformation. For instance, HKUST-1 has been converted to MIL-100 by the immersion of the former in a $FeCl_3·6H_2O$ solution for several hours at room temperature.[26] This is a clear example of a facile and mild transmetalation methodology to obtain MIL-100, which is conventionally produced by using harsh synthetic conditions (hydrofluoric acid, 150 °C, 6 days).[26] Another stimulating example is given by the transformation of the bio-MOF-1 to a Ag-based MOF by exposing the former to a silver nitrate solution.[27] The transformed Ag-MOF displayed an intense light blue emission, not observed in the pristine bio-MOF-1, thus conferring the fascinating properties of this new material.

Inspired by the foregoing examples, we investigate the MOF-to-MOF transformation from a non-luminescent material to highly luminescent MOF systems. The transmetalation processes described above require the extended immersion of the MOF in the desired metal salt solution from several hours to weeks, making this methodology sluggish and less practical.

Herein, we report a facile and fast methodology, taking advantage of the uncoordinated oxygen atoms of the terephthalate linkers present in the OX-1 MOF.[28-30] These uncoordinated oxygen atoms will promptly interact with the Ag cations of a silver nitrate ($AgNO_3$) solution, leading to the transformation of OX-1 (Zn-BDC, where BDC = 1,4-benzene dicarboxylate) to OX-2 (Ag-BDC) MOF, termed as Ag/OX-1 materials. Remarkably, and unlike the pristine OX-1, the new Ag/OX-1 transformed materials exhibit an intense green luminescence with an emission quantum yield of up to 22% in the solid-state (powder) form. Additionally, this green emission is strongly quenched in the presence of acetone (>90% reduction in initial emission intensity), turning these materials into attractive candidates for the future fabrication of breathalyzers for diabetes mellitus detection and monitoring.[31-33] This transformation mechanism was also extended to lead (Pb) cations. As they are well-known toxic pollutants present in aquatic environments,[34, 35] a fast conversion of OX-1 to Pb-BDC MOFs could be an easy to implement low-cost route to capture Pb ions from polluted water. To this end, we dispersed the OX-1 MOF in a hydrophobic polyurethane (PU) to fabricate a mixed-matrix membrane, which was then tested by immersing it in a $Pb(NO_3)_2$ water solution. After just 20 minutes of immersion, we proved that OX-1 is effectively transformed to a Pb-BDC MOF, and therefore this proof-of-concept may be exploited for the depuration of aquatic environments.

**Results and Discussion**

**Fast Transformation of OX-1 to OX-2 in Methanol**

To begin with, we explore the post-synthetic MOF transformation by leveraging the uncoordinated oxygen atoms of the BDC linker found in the OX-1 MOF (**Fig. 1A**). The Ag cations of a solution of $AgNO_3$ in methanol will promptly interact with the uncoordinated oxygen atoms, leading to the transformation of the non-luminescent OX-1 to the luminescent OX-2 MOF. To unveil the mechanism of the MOF transformation, the crystalline structure, morphology and spectroscopic properties were determined by means of powder X-ray diffraction (PXRD), Fourier transform infrared (FTIR) spectroscopy, scanning electron microscopy coupled to energy dispersive X-ray spectroscopy (FESEM-EDX), atomic force microscopy (AFM) coupled to local probe nanoFTIR, and steady-state fluorescence spectroscopy.

Upon the immersion of 250 mg of OX-1 MOF in a solution of 120 mg (0.7 mmol) of AgNO$_3$ in 20 mL of MeOH (see experimental part), an instantaneous change from a non-luminescent to a highly luminescent material exhibiting a green emission was observed under UV (365 nm) lamp irradiation (see video demo in Supporting Information, SI). Four different batches were prepared by interrupting the transformation reaction at different times ranging from 30 s to 24 hr, yielding the materials termed as Ag/OX-1 30s, Ag/OX-1 5min, Ag/OX-1 1 hr and Ag/OX-1 24 hr. Fig. 1B shows the PXRD patterns of the pristine OX-1 and OX-2 MOFs along with the transformed Ag/OX-1 materials. While the Bragg diffraction peaks of the pristine OX-1 and OX-2 resemble those previously reported,[28, 30, 36] Ag/OX-1 materials reveal a combination of diffraction peaks characteristic of the OX-1 and OX-2 MOFs, even after just 30 s of reaction (Fig. 1B). For the Ag/OX-1 materials converted at times longer than 30 s, the Bragg peak characteristic of OX-1 at a 2$\theta$ angle of ~5° (plane (200)) completely vanishes, indicating that at least part of the OX-1 MOF crystals is transformed to OX-2. However, after 24 h of reaction, there are still some diffraction peaks typical of OX-1 (Fig. 1B), which leads to the conclusion that even after this period, there is a partial, but not complete MOF-to-MOF transformation. In the same line, the FTIR spectra provide more proof about the conversion (Fig. 1C). While most of the peaks at lower energies remain unaffected, as they are characteristic vibrations of the BDC linker,[29, 37, 38] the peaks at ~1600 and ~1390 cm$^{-1}$, associated with the asymmetric and symmetric stretching vibrations of the carboxylic groups of the BDC linker coordinated to the metal centre[29, 39] are strongly shifted to lower energies (~1520 and 1360 cm$^{-1}$, Fig. 1C). The observed shifts agree with a change in the coordination of the BDC linkers, which will be mostly linked to the Ag ions instead of the Zn ones. This interpretation is confirmed by the fact that the FTIR spectrum of pristine OX-2 MOF exhibits similar IR vibrations at 1520 and 1360 cm$^{-1}$ assigned to the asymmetric and symmetric stretching vibrations of the carboxylic groups of the BDC linkers.[40] However, the contribution of the bands at 1600 and 1390 cm$^{-1}$ is still being detected (Fig. 1C), revealing that not all OX-1 crystals are fully converted to OX-2, as previously deduced from the XRD results. The rapid transformation of OX-1 to OX-2 supports our theory that the uncoordinated oxygen atoms of BDC linkers are responsible for this transmetalation reaction. To interrogate this effect further, we applied the same methodology but by replacing OX-1 with MOF-5, which is built using the same Zn cations and BDC linkers, however, the oxygen atoms in the latter framework are fully coordinated (Fig. 1D inset).[41, 42] The XRD pattern of the MOF-5 after being soaked for 24 hr in a methanolic solution of $AgNO_3$ is the same as the pristine MOF-5 (Fig. 1D). In addition, no green light emission was observed under UV irradiation. This result unequivocally proves that the transformation does not occur for MOF-5, evidencing that the reaction in OX-1 may involve the uncoordinated oxygen atoms in the linkers instead of the commonly described mechanism that exploits the uncoordinated metal sites.[13]

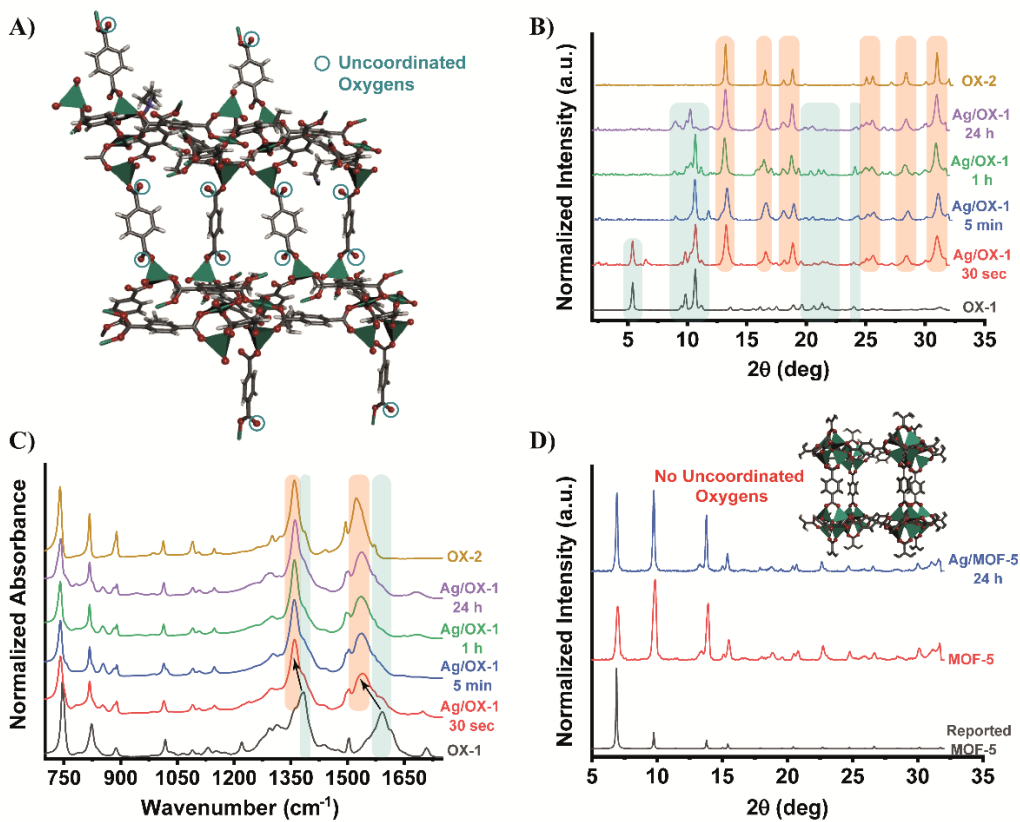

**Figure 1. A)** Representation of the 3D structure of OX-1 MOF, where the blue circles indicate the uncoordinated oxygen atoms. **B)** PXRD patterns and **C)** FTIR spectra of pristine OX-1 and OX-2 MOFs and the different converted Ag/OX-1 materials. **D)** PXRD patterns of the reported and as-synthesized MOF-5 and the non-transformed Ag/MOF-5 material. The inset is a representation of the 3D structure of MOF-5.

To unravel how the OX-1 conversion affects the morphology of the crystals, and to further answer the question about the metal composition of the Ag/OX-1 composite, we have performed the FESEM-EDX characterization (Fig. 2 and Fig. S1-4 in SI). Interestingly, the SEM micrographs of pristine OX-1 show a heterogeneous distribution of micron-sized crystals (Fig. S1A) as well as small elongated nanoplates (Fig. S1B). This heterogeneous distribution will be crucial, as the large surface area of the small nanoplates can facilitate the transformation process. This hypothesis is confirmed by carefully examining the FESEM micrographs and the corresponding EDX maps of the transformed Ag/OX-1 materials. Fig. 2 illustrates different regions of the Ag/OX-1 30 s. In these images, we pinpointed two distinctive regions: (i) the larger micro-sized crystals which are mainly composed of Zn metal (Fig. 2A-F), and (ii) relatively smaller nano-sized crystals, where only Ag traces can be determined (Fig. 2G-H). We can thus conclude that the large crystals of OX-1 remain stable when immersed in the methanolic Ag solution, whereas the nanoplates – with their larger surface area – transform from OX-1 to OX-2, as the Zn ions are fully replaced by Ag ions.

The coexistence of the larger crystals of OX-1 and the finer nanostructures of OX-2 is consistent with the results obtained from the PXRD and FTIR studies. FESEM micrographs and EDX maps of the transformed materials at different times of reactions (Ag/OX-1 5 min, 1 hr and 24 hr) also reveal the coexistence of the OX-1 large crystals and the OX-2 nanostructures (transformed from OX-1), as shown in Fig. S2-S4. To further validate these results, the crystals of different sizes have been investigated by AFM coupled with near-field infrared spectroscopy (nanoFTIR). We have recently demonstrated that this technique is a powerful tool to chemically characterize MOFs at a single crystal level with a resolution down to 20 nm.[43] Fig. 3 shows the AFM topography images belonging to the large (Fig. 3A) and small (Fig. 3B) crystals found in the sample after the transformation process, together with the nanoFTIR spectra locally probed on each type of crystals (Fig. 3C). The nanoFTIR spectrum of the large crystals (region 1) resembles that of OX-1 with bands at ~1600 and ~1390 cm$^{-1}$, while the spectrum recorded on the small nanocrystals (region 2) is similar to that observed for OX-2 crystals with characteristics peaks at ~1520 and ~1360 cm$^{-1}$. These findings unambiguously prove that the larger micron-sized crystals correspond to the non-transformed OX-1 MOF, while the smaller nanocrystals are OX-2 MOF nanoplates transformed from OX-1.

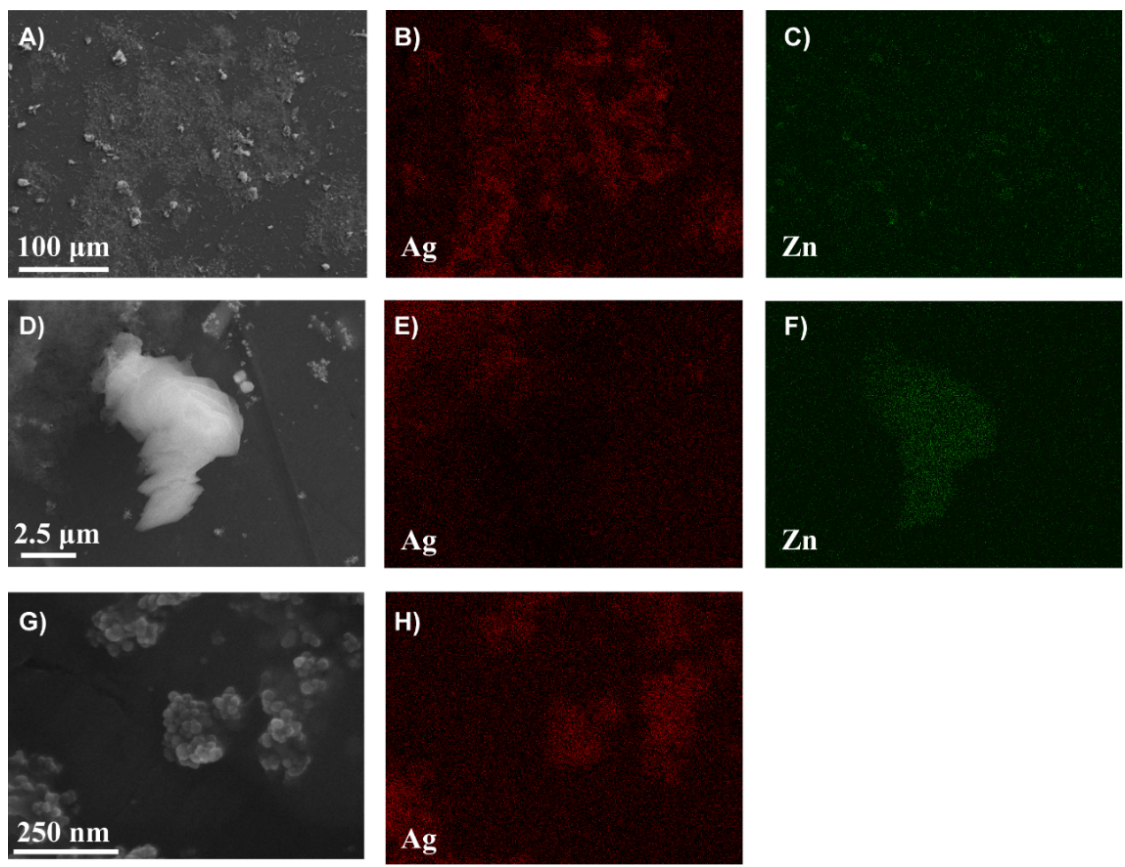

**Figure 2.** FESEM-EDX images of different regions found in Ag/OX-1 30 s. **A-C)** Display a large area where big OX-1 and small OX-2 crystals coexist. **D-E)** A larger crystal of OX-1 where only Zn atoms are detected, while **G-H)** is a magnified region containing only OX-2 (Ag-based) small crystals.

The main objective behind this MOF-to-MOF transformation was to impart a fluorescent property on the new transformed MOFs, as we previously reported that OX-2 is highly luminescent.[40] The transformed Ag/OX-1 (30 s, 5 min, 1 hr and 24 hr) materials are all highly emissive with a fluorescence quantum yield of around ~20% in the solid-state powder form (Fig. 4A,C and S5). The emission spectra resemble that of pristine OX-2 MOF, in which the photoexcitation at 315-340 nm induces an intense emission with a well-resolved vibrational structure having maxima at 480, 520 and 560 nm.[40] This emission was previously assigned to a ligand-to-metal charge transfer (LMCT) phenomenon.[40] However, the luminescent quantum yield of pristine OX-2 was higher (45%) than that of the transformed Ag/OX-1 material, suggesting that the OX-1 large crystals are absorbing part of the irradiating light and therefore, diminishing the quantum yield (as they are not luminescent). Interestingly, we found that the emission of these new materials is very sensitive to the presence of acetone. While the emission of Ag/OX-1 exhibits smaller changes in many different solvents, its emission almost disappears in the presence of acetone (~90% quenched emission), as shown in Fig. 4B and 4D. These are very relevant results as the detection of acetone in human breath is a highly promising alternative to commercially available methods for the diagnosis and monitoring of type 1 diabetes mellitus (such as urine or blood tests), which are more invasive, painful and could be unsafe.[33] It is well-known that when the level of insulin is insufficient, the human body cannot convert the glucose into energy and instead, the cells burn fats, generating acetone during the process.[32, 33] The levels of acetone for a healthy individual ranges from 0.2-1.8 ppm while patients suffering a diabetes episode can exhale from 1.25 to 25 ppm of acetone.[31, 32] Therefore, the ability to detect small amounts of acetone is paramount, stimulating the development of efficient sensors to enable the accurate detection and precise management of diabetes mellitus. Diverse methods have been proposed, however most of them are expensive, non-portable, or require operation in high temperatures (e.g. oxide-based sensors).[33, 44, 45] An alternative attracting much attention is based on fluorometric sensors,[30, 46-48] which can be of lower cost, portable, highly sensitive, and present a faster response.[3] Although more tests must be done (e.g., vapour sensing studies), the selective response of Ag/OX-1 materials towards acetone positions them as a promising candidate for deployment as a fluorometric acetone sensor.

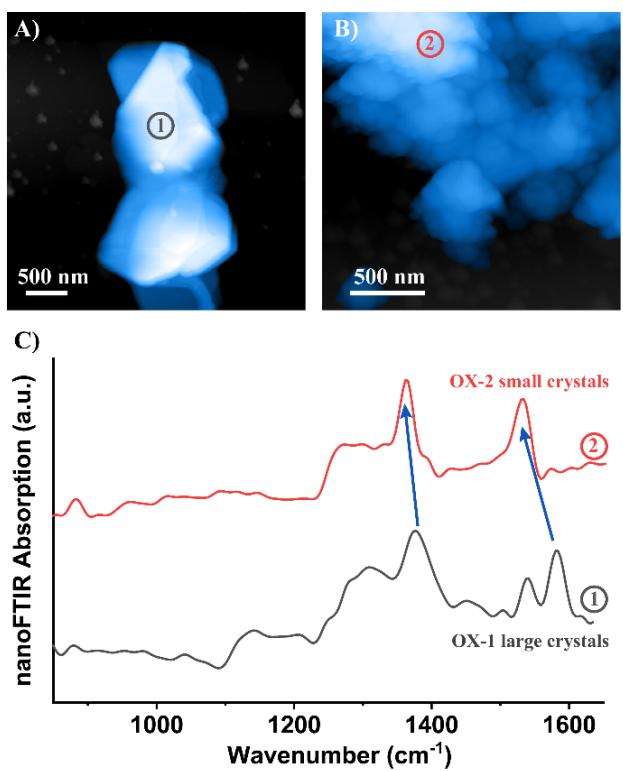

**Figure 3.** AFM topography images of Ag/OX-1 24h at selected regions: **A)** large crystals, and **B)** small nanocrystals. **C)** Local probe nanoFTIR spectra of Ag/OX-1 24 h recorded at the regions shown in panels (A) and (B), respectively.

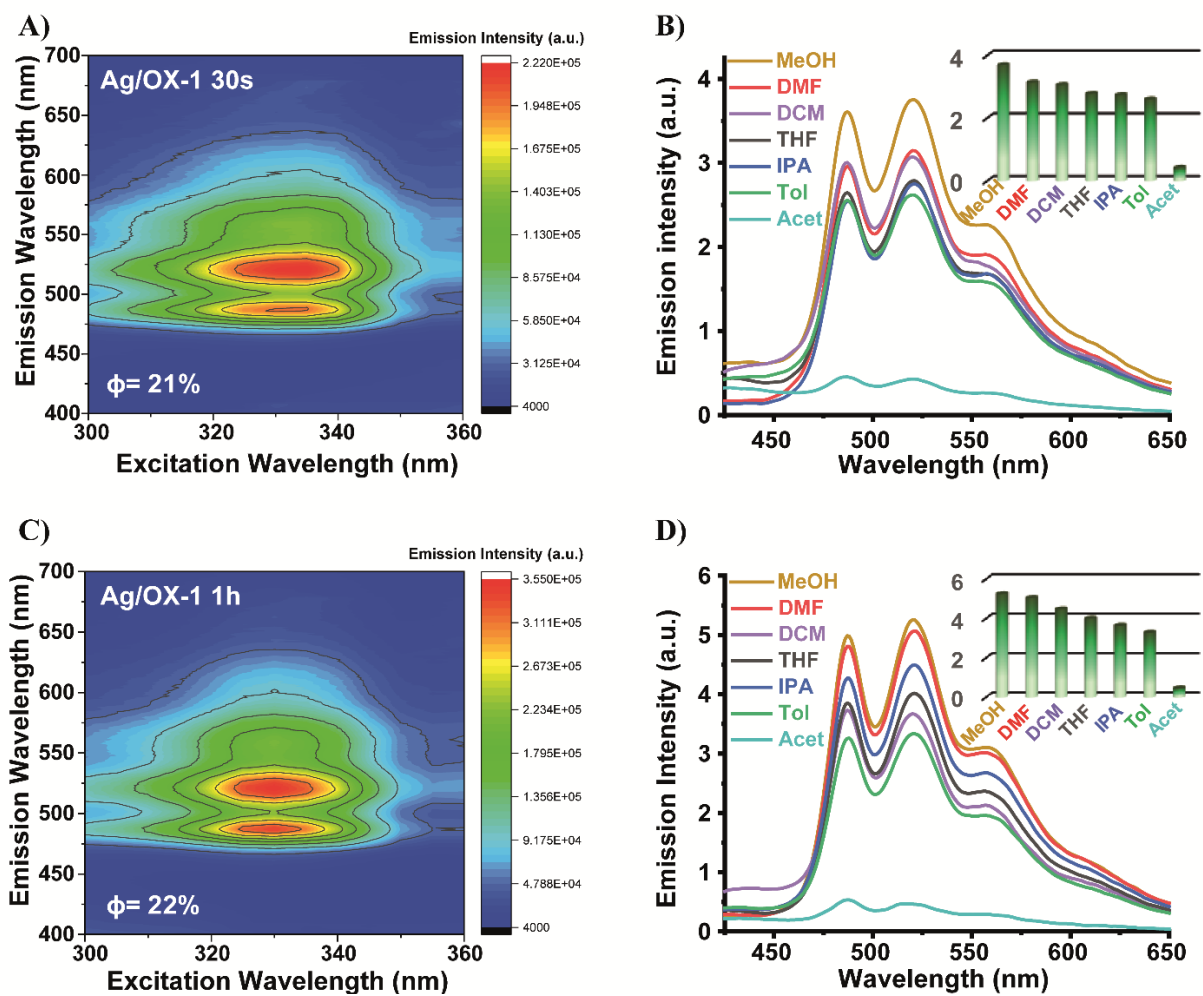

**Figure 4.** Excitation-emission maps of **A)** Ag/OX-1 30 s and **C)** Ag/OX-1 1 hr. The inset shows the emission quantum yield of each sample upon excitation at 330 nm. Emission spectra of **B)** Ag/OX-1 30 s and **D)** Ag/OX-1 1 hr in the presence of different VOCs. The inset is a graphical representation of the emission intensity maximum of Ag/OX-1 materials in each solvent, reflecting the significant quenching detected in acetone. The samples were excited at 330 nm.

**One-pot synthesis of OX-1 and OX-2 mixture**

To investigate whether the materials derived from the transformation of OX-1 to OX-2 is distinct from a direct production, we have synthesized a mixture of OX-1 and OX-2 in a one-pot reaction (see experimental part). SEM micrographs and the corresponding EDX maps reveal the co-existence of both OX-1 and OX-2 crystals (**Fig. S6A-E**). Similar to the Ag/OX-1 materials, in the one-pot synthesis, the OX-1 crystals are relatively larger than the OX-2 crystals. In addition, however, important structural changes were observed. Contrary to the transformed Ag/OX-1 materials, where the Bragg peak of OX-1 at $2\theta$ angle of 5° vanishes (suggesting the loss of some long-range periodicity), this peak contributes strongly to the signal in the mixture obtained through a one-pot synthesis (**Fig. S6F**). Moreover, all the diffraction peaks corresponding to OX-1 and OX-2 crystalline structures are identified in the PXRD pattern (**Fig. S6F**). This superposition suggests the coexistence of two-well differentiated phases of OX-1 and OX-2 crystals.

**Fig. S6G** displays the emission map of the material mixture, where, upon excitation at 315-340 nm, a well-resolved vibrational emission band with maxima at 480, 520 and 560 nm is observed. Interestingly, the quantum yield is now 45% which is exactly the same value as for the pristine OX-2 MOF synthesized in methanol.[40] This supports the fact that in the one-pot reaction, two distinguishable phases of OX-1 and OX-2 MOFs were obtained, as opposed to the one obtained through the transformation. This also affects the ability to detect acetone. While the transformed Ag/OX-1 materials are very sensitive to the presence of acetone, the mixture does not achieve the same sensing capabilities (**Fig. S6H-I**). Although the emission of the related mixture is quenched in the presence of acetone, it is just reduced by 60% (compared to the highest value), very far from the 90% of quenching observed for the Ag/OX-1 transformed materials. Additionally, the emission quenching of the mixture in the presence of acetone is not very different from the one observed for toluene or DCM (**Fig. S6H-I**), indicating that this mixture is much less selective towards acetone.

**Environmentally Friendly Transformations of OX-1 to OX-2**

With a view of fabricating the transformed Ag/OX-1 materials through an eco-friendlier route, we have explored new alternatives like a mechanochemical transformation or the use of deionized water instead of methanol. The transformation of OX-1 to OX-2 in deionized water was performed by mixing powders of OX-1 (250 mg) and $AgNO_3$ (120 mg, 0.7 mmol) followed by the addition of small amounts of deionized water (50-250 µL) and a brief mixing of all the reactants (see experimental part). As OX-1 is not very stable in water, we applied the same methodology to pristine OX-1 (addition of 250 µL of deionized water to 250 mg of OX-1) to test whether OX-1 degrades under these conditions. As shown in **Fig. S7A**, the PXRD spectrum of OX-1 after the addition of 250 µL water maintains the main characteristic Bragg diffraction peaks, indicating that no important structural changes occur upon the addition of such small amounts of water. On the other hand, in the presence of $AgNO_3$ and water, the OX-1 is transformed to OX-2 in a similar way to that described in methanol. The PXRD spectra show a combination of OX-1 and OX-2 diffraction peaks, but the OX-1 peak at $2\theta$ angle of 5° disappears (**Fig. S7A**). The FTIR spectra (**Fig. S7B**) yield very similar results to those described above, where the bands in the lower energy region remain unaffected, while the bands at ~1600 and ~1390 $cm^{-1}$ are strongly shifted to lower energies, agreeing with a coordination between BDC and Ag. Similar results were observed for the materials obtained through the mechanochemical OX-1 to OX-2 transformation. In this case, 250 mg of OX-1 and 120 mg of $AgNO_3$ were ground by mortar and pestle. For one batch, the powders were ground for 30 min, while in a second batch, another mixture was ground for 2 min after the addition of one drop of water in order to facilitate the reaction (see experimental part). In both cases, the PXRD spectra show the combination of OX-1 and OX-2 diffraction peaks, along with the disappearance of the OX-1 peak at 2$\theta$ angle of 5° (**Fig. S8A**), and the FTIR spectra illustrate the shift of the bands at ~1600 and ~1390 cm$^{-1}$ to lower energies (**Fig. S8B**).

As expected, the transformed Ag/OX-1 materials exhibit a green luminescent emission. As an example, **Fig. S8C-D** displays a photo of the transformed materials upon UV (365 nm) irradiation. First, the OX-1 MOF is not emissive in the visible spectral range, but during the grinding process, the mixture becomes luminescent due to the rapid transformation of OX-1 to OX-2 (**Fig. S8C**). In a similar way, the addition of 1 drop of water produces an instantaneous reaction, which facilitates the mechanochemical transformation process (**Fig. S8D**). The excitation-emission maps of the Ag/OX-1 transformed materials in water are comparable to the ones of the materials synthesized in methanol (**Fig. S9**). Interestingly, the fluorescence quantum yield value increases with the amount of water used (from 12 to 16%), presumably because more OX-1 is being transformed into OX-2. The transformed Ag/OX-1 (250 µL H$_2$O) material exhibits an astonishing response to acetone. In the presence of this volatile organic compound (VOC), the emission intensity is quenched by almost 100%, while it is barely affected by the presence of other VOCs (**Fig. S9D**). The materials obtained through the mechanochemical procedure unveil excitation-emission maps (**Fig. S10A-B**) that are comparable to those of **Fig. 4A,C**. However, the emission quantum yield is relatively lower (3-5%), indicating that the mechanochemical transformation is less effective. The excellent ability of this Ag/OX-1 material to selectively detect acetone is presented in **Fig. S10C-D**. Based on the above results, we not only demonstrate the possibility of transforming OX-1 to OX-2 by employing a green methodology, but more importantly, we unravel that the obtained materials could become promising candidates for developing acetone sensors.

**Transformation of OX-1 to Pb-MOF**

To exploit and further demonstrate the versatility of the MOF transformation mechanism described above, we explored the conversion of OX-1 to other type of MOF incorporating a different metal ion. We selected Pb as the secondary metal ion based on two main reasons: (i) firstly because it has been previously reported that Pb-BDC MOFs are luminescent,[49] and thus, the transformation could be easily followed by irradiating the mixture with a UV light, and (ii) secondly and most importantly, Pb metal ions are a highly toxic pollutant that can be found in aquatic environments.[34, 35] Therefore, if OX-1 can be quickly transformed to a Pb-MOF, it could be an interesting approach to capture Pb ions from polluted water.

We followed the same methodology described above for the Ag/OX-1 materials, but now 250 mg of OX 1 were soaked in a solution of 165 mg (0.5 mmol) of $Pb(NO_3)_2$ in 20 mL of MeOH for 20 min (see experimental part). **Fig. 5A** shows the PXRD patterns of pristine OX-1 and Pb-BDC MOFs and the Pb/OX-1 transformed material. Remarkably, the PXRD spectrum of the transformed Pb/OX-1 matches that of the pristine Pb-BDC MOF,[50-52] and in contrary to the Ag/OX-1 materials, no features of OX-1 diffraction Bragg peaks were observed. This fact implies a complete transformation of OX-1 to the Pb-BDC MOF in the first 20 min, suggesting a higher affinity towards Pb ions than to the Ag ones. As expected, this transformation produces a luminescent Pb-BDC MOF (quantum yield of 8%), whose emission is characterized by a broad band with an intensity maximum at around 500 nm upon excitation at 315-350 nm (**Fig. 5B**).

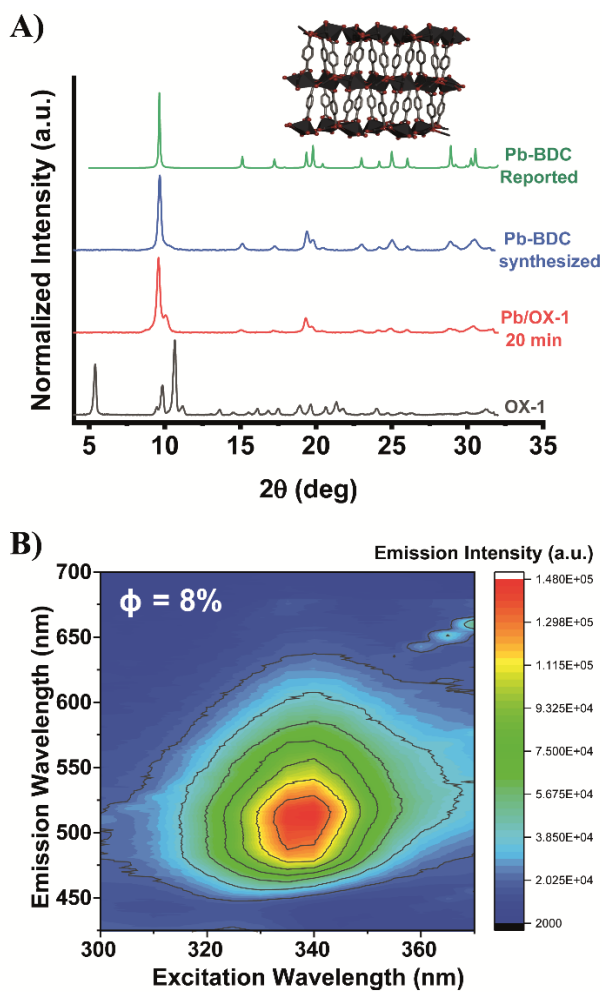

**Figure 5. A)** PXRD patterns of pristine OX-1 and Pb-BDC (as-synthesized and reported) MOFs and the converted Pb/OX-1 material. **B)** Excitation-emission map of Pb/OX-1 material. The inset is the emission quantum yield of the sample upon excitation at 330 nm.

The extremely fast and complete transformation of OX-1 to Pb-BDC MOF encourages us to exploit this mechanism for the capturing of Pb ions in water. However, some important aspects must be considered prior to that. Firstly, the synthesis of Pb-BDC MOF in water generates a different MOF structure,[53] and therefore the OX-1 conversion could be different. Secondly and more crucially, OX-1 MOF degrades when immersed in water, hindering its possible applicability for Pb trapping from aquatic media. To overcome the latter issue, we have dispersed the OX-1 MOF in a hydrophobic polyurethane membrane (OX-1/PU see experimental part). The fabrication of mixed-matrix membranes (MMM) composed of a MOF and a polymer is a well-established procedure for boosting the use of MOFs in disruptive technologies in applications such as gas or liquid separation or water purification.[54-58] The idea behind this is to synergistically combine the excellent properties of MOFs (high surface area, long range periodicity, etc.) and polymers (ease of processing and tunable mechanical properties).[59-61] In our case, we envisaged that the polyurethane membrane (due to its rubbery nature)[62] will permit a controlled diffusion of the Pb water solution, thus preventing the degradation of OX-1 MOF. This assumption was further corroborated by PXRD experiments (**Fig. S11**), which showed that the crystalline structure of OX-1 in PU remains stable after its immersion in water for 20 mins (duration of the experiment). In contrast, we note that the pristine OX-1 instantaneously degrades in water (when not protected within PU matrix).

Once we established that OX-1/PU MMM is resilient enough to water exposure, we investigated whether the OX-1 in the composite membrane can be transformed to Pb-BDC MOF under these conditions, and therefore, whether the OX-1/PU membrane is capable of capturing Pb ions from a water solution. To proof this concept, a film of OX-1/PU (size: ~2.5 x 2.5 cm$^2$) was immersed in a solution of Pb(NO$_3$)$_2$ in water (0.05 M) without stirring for 20 mins. The film was then dried at room temperature; and the conversion of OX-1 to Pb-BDC MOF was confirmed by PXRD and luminescence experiments. Fig. 6A shows the PXRD spectra of OX-1/PU, pristine Pb-BDC-H$_2$O (synthesized in water) and the transformed Pb-OX-1/PU. The PXRD pattern of the latter exhibits the Bragg diffraction peaks characteristic of Pb-BDC-H$_2$O,[53] indicating a conversion from OX-1 to Pb-BDC MOF. The OX-1 to Pb-BDC conversion was also confirmed by the turquoise luminescence of the Pb-OX-1/PU membrane. Fig. 6B displays the photos of the as-synthesized OX-1/PU film prior to and after its interaction with Pb ions under UV (365 nm) lamp irradiation. The pristine OX-1/PU membrane exhibits a light blue emission typical of the PU membrane. On the contrary, the Pb-OX-1/PU membrane unveils a turquoise emission, whose intensity is gradually increasing with the time exposed to the Pb solution. Fig. 6C shows the emission spectra of the pristine OX-1/PU and the converted Pb-OX-1/PU membranes. The emission spectrum of the former, attributed to the PU polymer, consists of a broad band with maximum located at

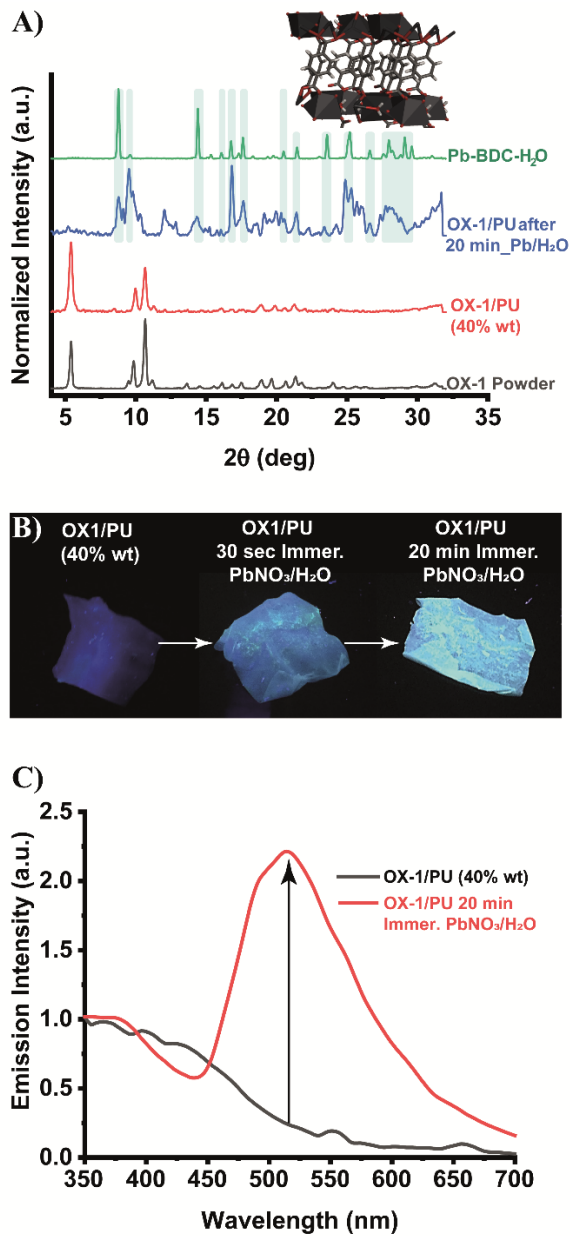

**Figure 6. A)** PXRD patterns of pristine OX-1 and Pb-BDC-H$_2$O MOF powders, and the OX-1/PU mixed-matrix membrane before and after immersing in a solution of Pb(NO$_3$)$_2$ for 20 minutes. **B)** Photo of the OX-1/PU membrane illuminated with an UV lamp (365 nm) before and after its exposure to a solution of Pb(NO$_3$)$_2$ for 30 seconds and 20 minutes, respectively. **C)** Emission spectra of the OX-1/PU membrane before and after its submersion in a solution of Pb(NO$_3$)$_2$ for 20 minutes. The samples were excited at 330 nm.

~375 nm. The emission spectrum of the Pb-OX-1/PU membrane depicts a similar emission band (at 375 nm) of the PU along with a main band at ~515 nm, corresponding to the emission of the converted Pb-BDC-H$_2$O MOF. Albeit preliminary, these results open an exciting pathway for the use of OX-1 mixed-matrix membranes to capture Pb ions from aquatic environments. Moreover, the fact that OX-1 can be converted to different MOFs using such a different cation such as Ag or Pb, leads us to believe that this methodology could be expanded to other metal cations or pollutants which can strongly interact with the uncoordinated oxygen atoms in the OX-1 MOF.

**Experimental Section**

**Materials**

Terephthalic acid (BDC, 99%+), zinc nitrate hexahydrate (Zn(NO$_3$)$_2$·6H$_2$O), zinc acetate dihydrate (Zn(OAc)$_2$·2H$_2$O), silver nitrate (AgNO$_3$), lead nitrate (Pb(NO$_3$)$_2$), triethylamine (NEt$_3$, 99%), *N, N*-dimethylformamide (DMF, 99%), dichloromethane (DCM, 99%+), acetone (Acet, 99.6% for spectroscopy), tetrahydrofuran (THF, 99.85%), isopropanol (IPA, 99.5%+), toluene (Tol, 99%+), and methanol (MeOH, 99.8%+) were purchased from Fisher Scientific and used without further purification.

**Synthesis of OX-1 MOF**

The OX-1 MOF was synthesized using the protocol described elsewhere with some modifications.[30] 12.0 mmol of BDC were deprotonated in a solution of 24.0 mmol of NEt$_3$ in 20 mL of methanol. A second solution was prepared by sonicating 6.0 mmol of Zn(NO$_3$)$_2$·6H$_2$O in 20 mL of methanol. The latter solution was quickly added to the former and a white suspension was promptly formed. The mixture was sonicated for 5 minutes and then the white solid sample was washed twice with methanol, collected by centrifugation (8000 rpm) and dried at 80 °C for 4 hours. This procedure yielded 2.5 g of OX-1 MOF.

**Conversion of OX-1 to OX-2 in methanol**

0.7 mmol of $AgNO_3$ were sonicated until the complete dissolution of the crystals in 20 mL of methanol. Then, 250 mg of OX-1 were poured into the above solution and the mixture was stirred for a certain time. Four different samples were prepared, stopping the reaction at 30 sec, 5 mins, 1 hr and 24 hr, respectively. Then the solvent was removed, and the sample was briefly washed once with methanol to remove the possible excess of Ag ions. The sample was finally dried at 80 °C for 4 hours.

**Synthesis of MOF-5**

MOF-5 was synthesized following the recipe previously reported with some minor modifications.[42] 15 mmol of BDC were dissolved in a solution of 30 mmol of $NEt_3$ in 200 mL of DMF. Another solution was prepared by dissolving 38 mmol of $Zn(OAc)_2 \cdot 2H_2O$ in 200 mL of DMF. The zinc salt solution was added to the BDC one and the mixture was stirred for 2.5 hr. The white solid sample was washed twice with DMF and DCM, collected by centrifugation (8000 rpm) and dried at 120 °C under reduced pressure for 4 hours. This procedure yielded ~2.5 g of MOF-5.

**One-pot synthesis of OX-1/OX-2 mixture**

For the one-pot synthesis of the OX-1/OX-2 mixture, we followed a very similar methodology as for the synthesis of OX-1. Briefly, 6.0 mmol of BDC were deprotonated in a solution of 12.0 mmol of $NEt_3$ in 20 mL of methanol. Another solution was prepared by sonicating 1.5 mmol of $Zn(NO_3)_2 \cdot 6H_2O$ and 1.5 mmol of $AgNO_3$ in 20 mL of methanol. The second solution containing the mixture of metal ions was quickly added to the BDC one and a white suspension was promptly formed. The mixture was sonicated for 10 mins and the white solid sample was washed twice with methanol, collected by centrifugation (8000 rpm) and dried at 80 °C for 4 hr.

**Green conversion of OX-1 to OX-2: Mechanochemistry and deionized water**

Mechanochemistry: 250 mg of OX-1 and 0.7 mmol of $AgNO_3$ were ground by mortar and pestle for 30 mins (the humidity of the lab was 40%). The conversion was followed by irradiating the sample with UV light (**Fig. S8C**). The white solid sample was then dried at 100 °C for 2 h. A similar procedure was followed but with the addition of 1 drop of water in order to accelerate the grinding process. In this case the mixture was only ground for two minutes and a similar degree of OX-1 to OX-2 conversion was observed. As in the above example, the sample was dried at 100 °C for 2 h.

Water methodology: 3 different samples were prepared by the addition of increasing amounts of deionized water (50, 100 and 250 μL) to a mixture of 250 mg of OX-1 and 0.7 mmol of $AgNO_3$. The reactants were mixed for 1 min using a spatula, and the white paste was dried at 100 °C for 2 hr.

**Conversion of OX-1 to Pb-BDC MOF in methanol**

This procedure was similar to the one described above. Briefly, 0.5 mmol of $Pb(NO_3)_2$ were sonicated in 20 ml of methanol. Then, 250 mg of OX-1 were poured in the former solution and the mixture was stirred for 20 minutes. After that, the solvent was removed, and the sample was washed once with methanol to remove any excess Pb ions. The sample was finally dried at 80 °C for 4 hr.

**Fabrication of the OX-1/PU mixed-matrix membrane (MMM)**

The OX-1/PU MMM was fabricated similarly to previous reports.[63, 64] Firstly, a PU solution (15% w/w) was prepared by dissolving poly [4,4'-methylenebis(phenyl isocyanate)-alt-1,4-butanediol/di(propylene glycol)/polyurethane] pellets (Sigma Aldrich) in THF by stirring the mixture for 48 hr until the complete dissolution of the polymer pellets. Subsequently, a certain amount of OX-1 was dispersed in a small amount of THF (1 mL) by sonicating and stirring the suspension. This OX-1 dispersion was then incorporated in the PU-THF solution to yield 40 wt.% of MOF. The wt.% of MOF in the PU matrix was determined using the following equation:

$$\mathbf{MOF}\ wt.\% = \left(\frac{m_{MOF}}{m_{MOF}+m_{PU}}\right) \times \mathbf{100}$$

where $m_{MOF}$ and $m_{PU}$ are the weights of OX-1 MOF (dispersed in THF) and PU (dissolved in THF), respectively.

The MMM was finally obtained by doctor blading the OX-1/PU mixture (in THF), following by the drying of the composite film at room temperature. The membrane has an original size of around 10 x 10 cm$^2$ and a thickness of ~0.5 mm. For the Pb capture experiments the membrane was cut to a size of ~2.5 x 2.5 cm$^2$.

**Materials characterization**

The crystalline structure, morphology and luminescent properties of the materials were characterized by a combination of powder X-ray (PXRD), Fourier transformed infrared (FTIR) and fluorescence spectroscopy and field emission scanning electron microscopy coupled to energy dispersive X-ray (FESEM-EDX) spectroscopy techniques. PXRD experiments were conducted in a Rigaku Miniflex diffractometer with a Cu Kα source (1.541 Å). The diffraction data were collected using 0.01° step size, 1° min$^{-1}$ and for 2θ angle ranging from 2° to 32°. FESEM images and the corresponding EDX spectra were obtained using the Carl Zeiss Merlin equipped with a field emission gun. Micrographs were attained under high vacuum with an accelerating voltage of 10 keV and in secondary electron imaging mode. FTIR spectra were recorded on a Nicolet iS10 spectrometer. The FTIR spectrum of each sample was collected 3 times and then averaged.

Near-field spectroscopy and AFM imaging were performed on a neaSNOM instrument (Neaspec GmbH) based on a tapping-mode AFM. Here, the platinum-coated tip (oscillation frequency $\Omega$ = 254 kHz) is used both to map the height topography and to probe the optical near-field upon illumination with a broadband mid-infrared laser source (Toptica, Germany). The detector signal is demodulated at higher harmonics ($n\Omega$) to suppress the interference of background contributions. Each spectrum is acquired by Fourier transform spectroscopy averaging over 10 individual interferograms with an integration time of 14 ms, probed with a spot size of 20 nm on individual crystals. All spectra are normalized by reference spectra obtained on the silicon substrate.

Steady-state fluorescence spectra, excitation-emission maps, and luminescence quantum yields were recorded using a FS5 spectrofluorometer (Edinburgh Instruments) equipped with different modules for each specific experiment (i.e. integrating sphere for quantum yield, or modules for liquid or solid-samples).

For the emission experiments of Ag/OX-1 in presence of different VOCs, 2 mg of the MOF material was added to 4 mL of each solvent. The mixture was sonicated for 15 min to disperse well the MOF crystals. All the experiments were performed at room temperature.

**Conclusions**

In this study, we have established a post-synthetic approach to accomplish MOF-to-MOF transformation by leveraging the uncoordinated oxygen positions present in the BDC linkers of the OX-1 MOF. The conversion reaction is facile and fast, requiring only the immersion of a certain amount of OX-1 MOF in a silver metal salt solution for a short period of time (~minutes). We also demonstrated that the MOF transformation can be extended to green methodologies involving dry media through a mechanochemical reaction (grinding mechanism), or simply through the use of water (instead of methanol). The obtained Ag/OX-1 materials exhibit an intense green emission, which is selectively quenched in the presence of acetone, making them promising candidates for the future development of advanced sensor devices for diabetes monitoring. The underlying mechanism for acetone sensing is currently unknown and it warrants further investigation. Our protocol was also exploited for the rapid fully conversion of OX-1 to different Pb-BDC MOFs. This finding motivated us to prepare a mixed-matrix membrane by dispersing the OX-1 in a rubbery polyurethane membrane which was then employed as a proof-of-concept for capturing Pb ions from a water solution. Our work highlights the use of alternative synthetic routes to impart desired qualities to MOFs like luminescence, or even to take advantage of these approaches for emerging applications such as acetone sensing or water purification technologies.

## AUTHOR INFORMATION

**Corresponding Author**

*Jin-Chong Tan.

E-mail: jin-chong.tan@eng.ox.ac.uk

**Funding Sources**

This work was supported by the ERC Consolidator Grant through the grant agreement 771575 (PROMOFS).

**Supporting Information**

Supporting Information includes (i) FESEM-EDX images of the OX-1, Ag/OX-1, and OX-1/OX-2 materials; (ii) the excitation-emission maps of Ag/OX-1 at 5min and 24h; (iii) PXRD patterns, excitation-emission maps, and luminescence response to different VOCs of the OX-1/OX-2 mixture; (iv) PXRD and FTIR spectra of Ag/OX-1 materials transformed using water; (v) PXRD and FTIR spectra together with photographic images of the mechanochemically transformed Ag/OX-1 materials; (vi) excitation-emission maps, and luminescence response to different VOCs of Ag/OX-1 (transformed mechanochemically) and using water, respectively; and (vii) PXRD patterns of OX-1/PU membranes before and after water immersion study.

**Acknowledgements**

We thank the Research Complex at Harwell (RCaH) for access to the Nicolet iS10 FTIR spectrometer. We thank Dr. Barbara Souza and Dr. Jennifer Holter for their help with FESEM-EDX characterization.

# Supporting Information

## for

## Facile and Fast Transformation of Non-Luminescent to Highly Luminescent MOFs: Acetone Sensing for Diabetes Diagnosis and Lead Capture from Polluted Water

*Mario Gutiérrez,[1] Annika F. Möslein,[1] and Jin-Chong Tan[1]\**

[1]Multifunctional Materials & Composites (MMC) Laboratory, Department of Engineering Science, University of Oxford, Parks Road, Oxford OX1 3PJ, United Kingdom.

\*Corresponding author: jin-chong.tan@eng.ox.ac.uk

**Figure S1.** FESEM images of different regions found in OX-1. **A)** Shows microsized OX-1 crystals while B) illustrates the small nanoplates.

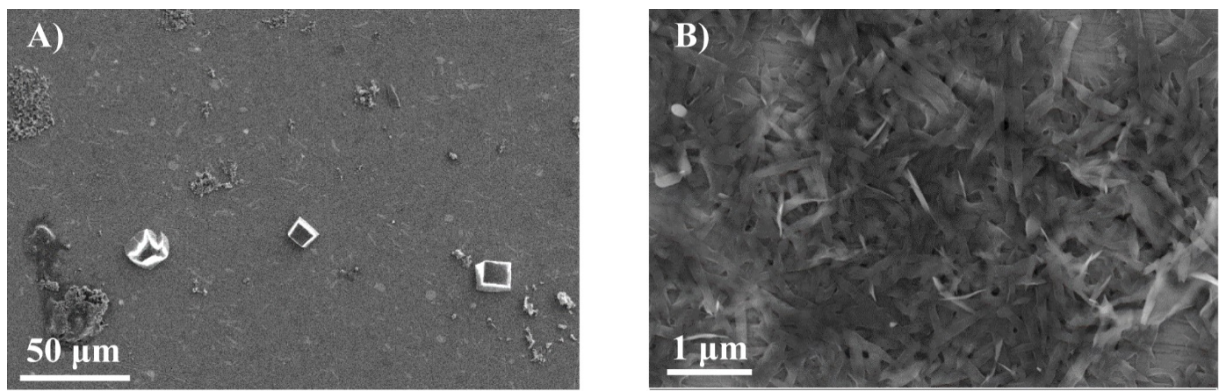

**Figure S2.** FESEM-EDX images of Ag/OX-1 5min showing the concurrent presence of OX-1 big crystals and small transformed OX-2 ones.

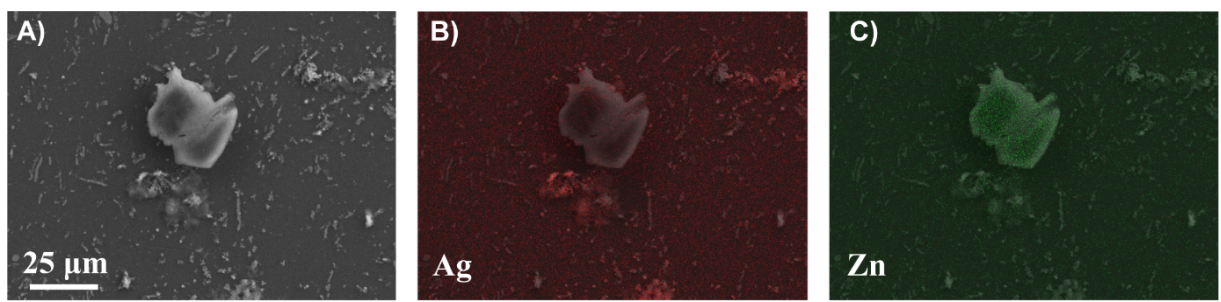

**Figure S3.** FESEM-EDX images of Ag/OX-1 1h showing the coexistence of big crystals of OX-1 and nanocrystals of transformed OX-2.

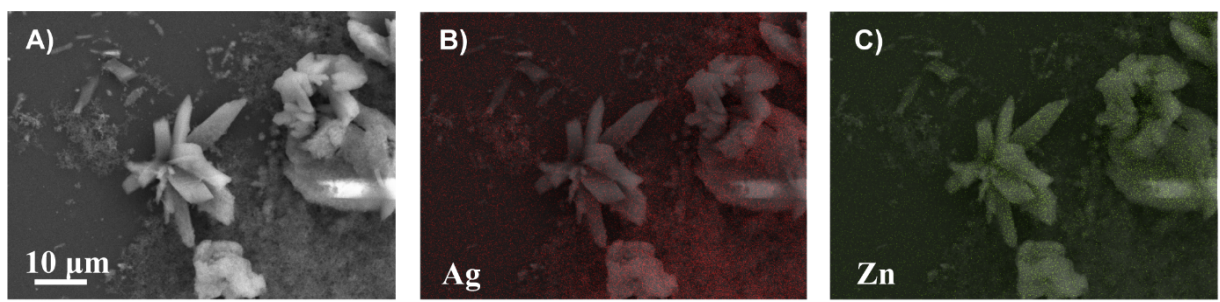

**Figure S4.** FESEM-EDX images of a large area found in Ag/OX-1 24h, where microsized OX-1 crystals as well as converted OX-2 nanocrystals are observed.

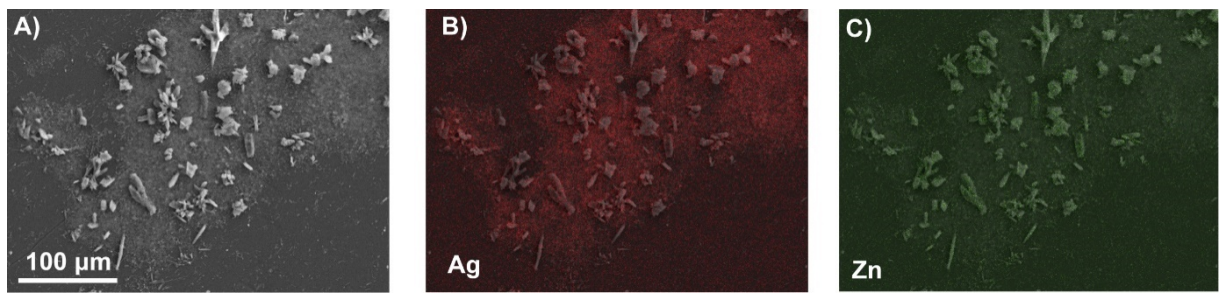

**Figure S5.** Excitation-emission maps of **A)** Ag/OX-1 5min and **B)** Ag/OX-1 24h. The inset shows the emission quantum yield of each sample upon excitation at 330 nm.

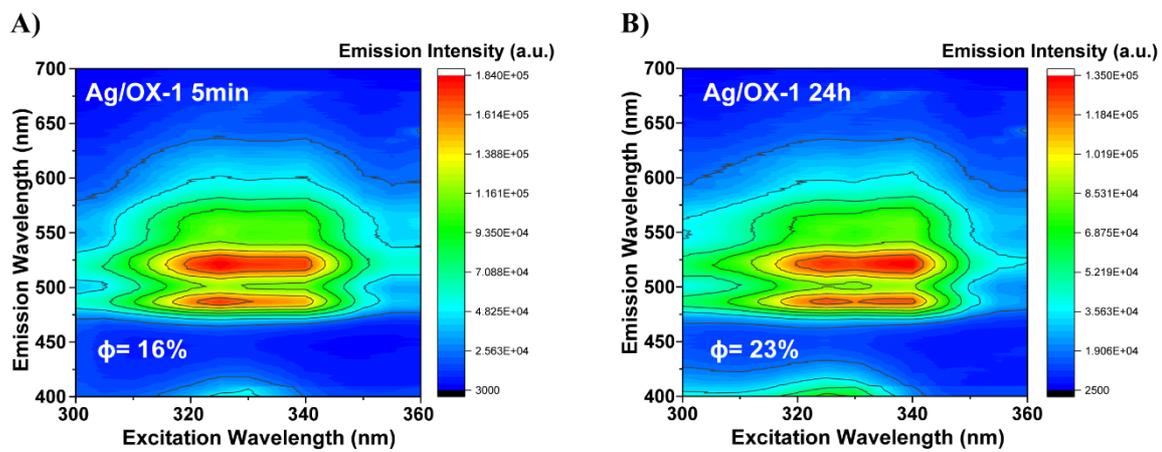

**Figure S6. A-E)** FESEM-EDX images of different regions found in OX-1/OX-2 mixture obtained through the one-pot synthesis method. **F)** PXRD patterns of pristine OX-1 and OX-2 MOF and the related OX-1/OX-2 mixture. **G)** Excitation-emission map of the OX-1/OX-2 mixture. The inset illustrates the emission quantum yield upon excitation at 330 nm. **H)** Emission spectra of the OX-1/OX-2 mixture in different solvents, and **I)** is a graphical representation of its emission intensity maximum in each solvent. The samples were excited at 330 nm.

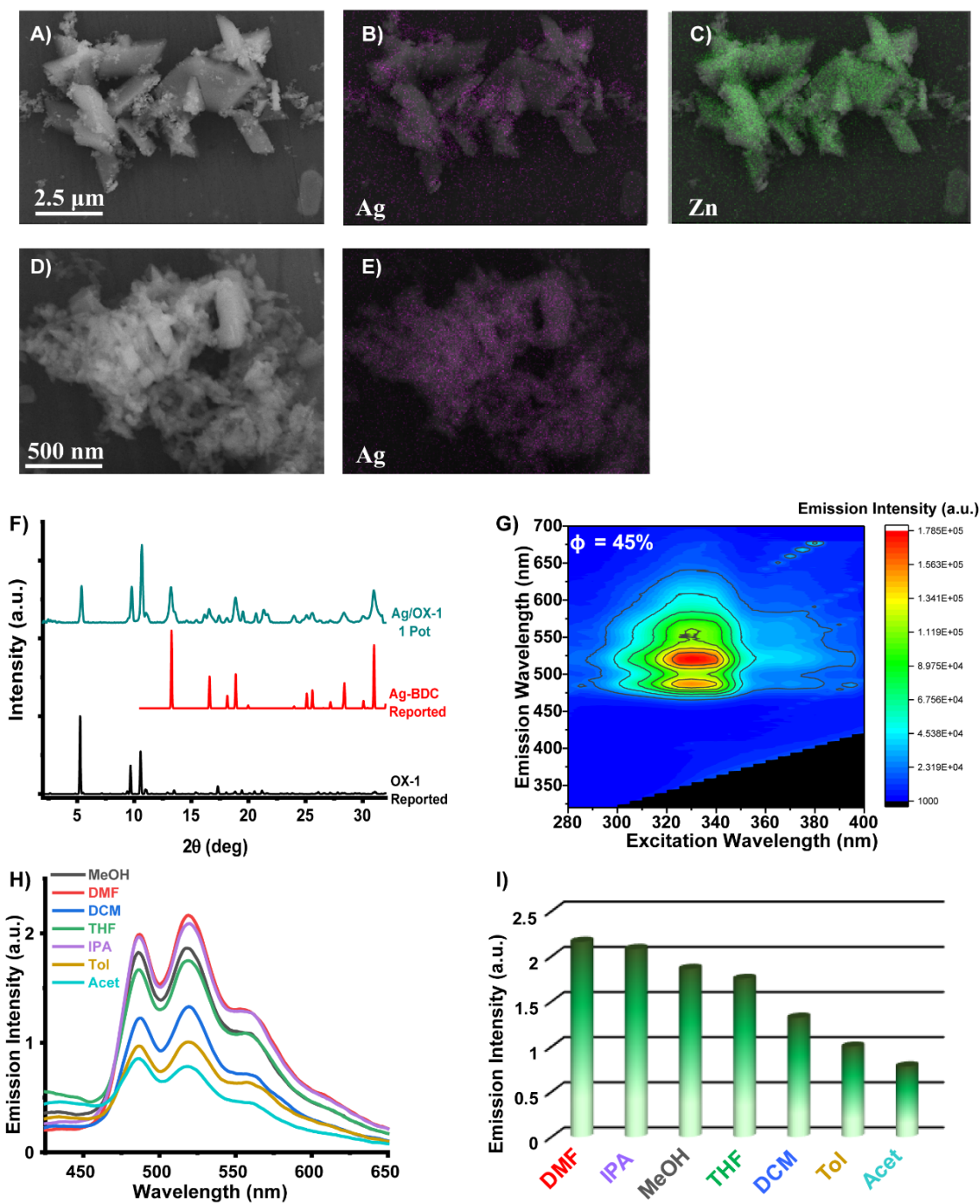

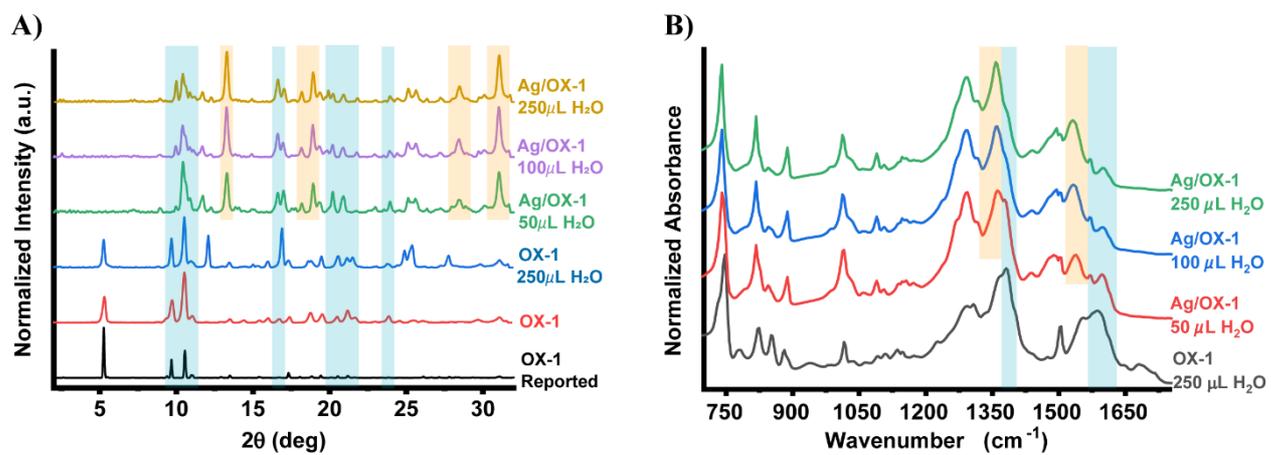

**Figure S7. A)** PXRD patterns and **B)** FTIR spectra of pristine OX-1, OX-1 treated with water and the converted Ag/OX-1 materials using different amounts of deionized water.

**Figure S8. A)** PXRD patterns and **B)** FTIR spectra of pristine OX-1, OX-1 treated with 1 drop of water and the mechanochemically converted Ag/OX-1. **C)** Photo of the OX-1 material before and after its conversion to OX-2 (grinding process) under UV irradiation. **D)** Photo of a mixture of AgNO$_3$ and OX-1 powders under UV irradiation and upon the addition of 1 drop of water and their subsequent grinding for 2 minutes.

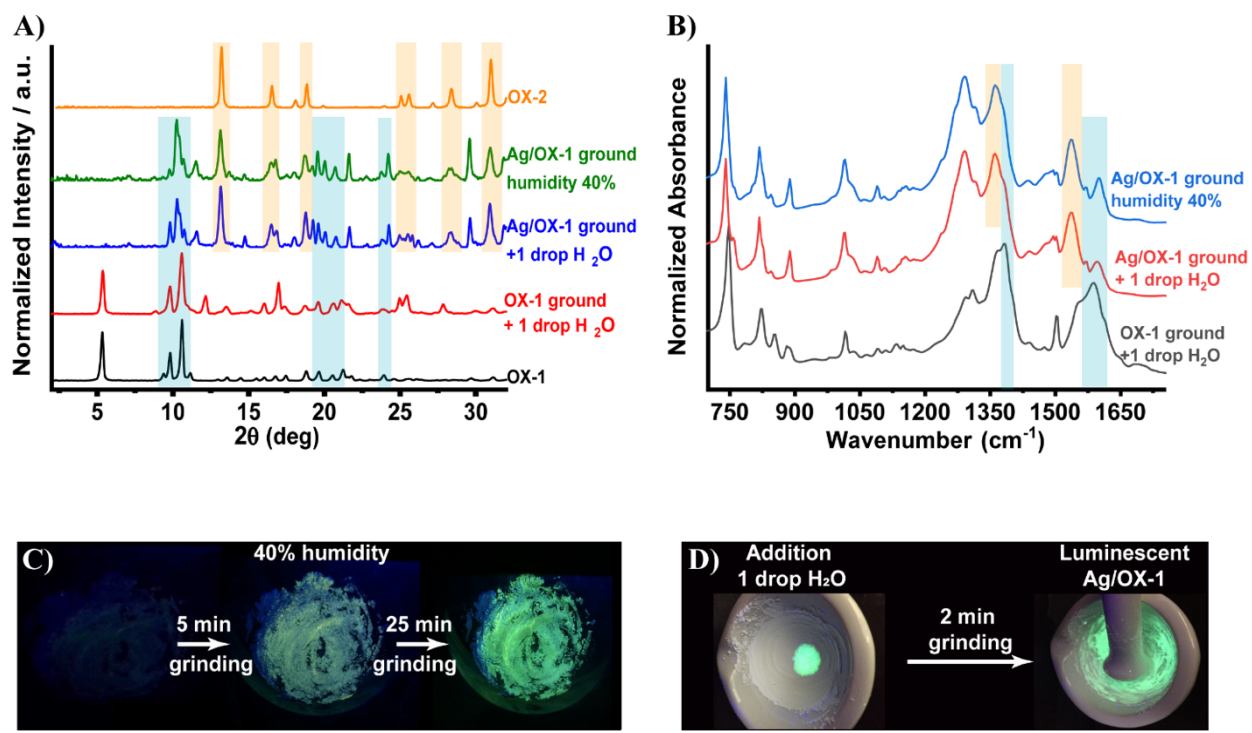

**Figure S9.** Excitation-emission map of **A)** Ag/OX-1 50μL H₂O, **B)** Ag/OX-1 100μL H₂O and **C)** Ag/OX-1 250μL H₂O materials. The inset shows a photo of the powders under UV irradiation and the emission quantum yield value of each sample upon excitation at 330 nm. **D)** Emission spectra of Ag/OX-1 250μL H₂O in different solvents. The samples were excited at 330 nm.

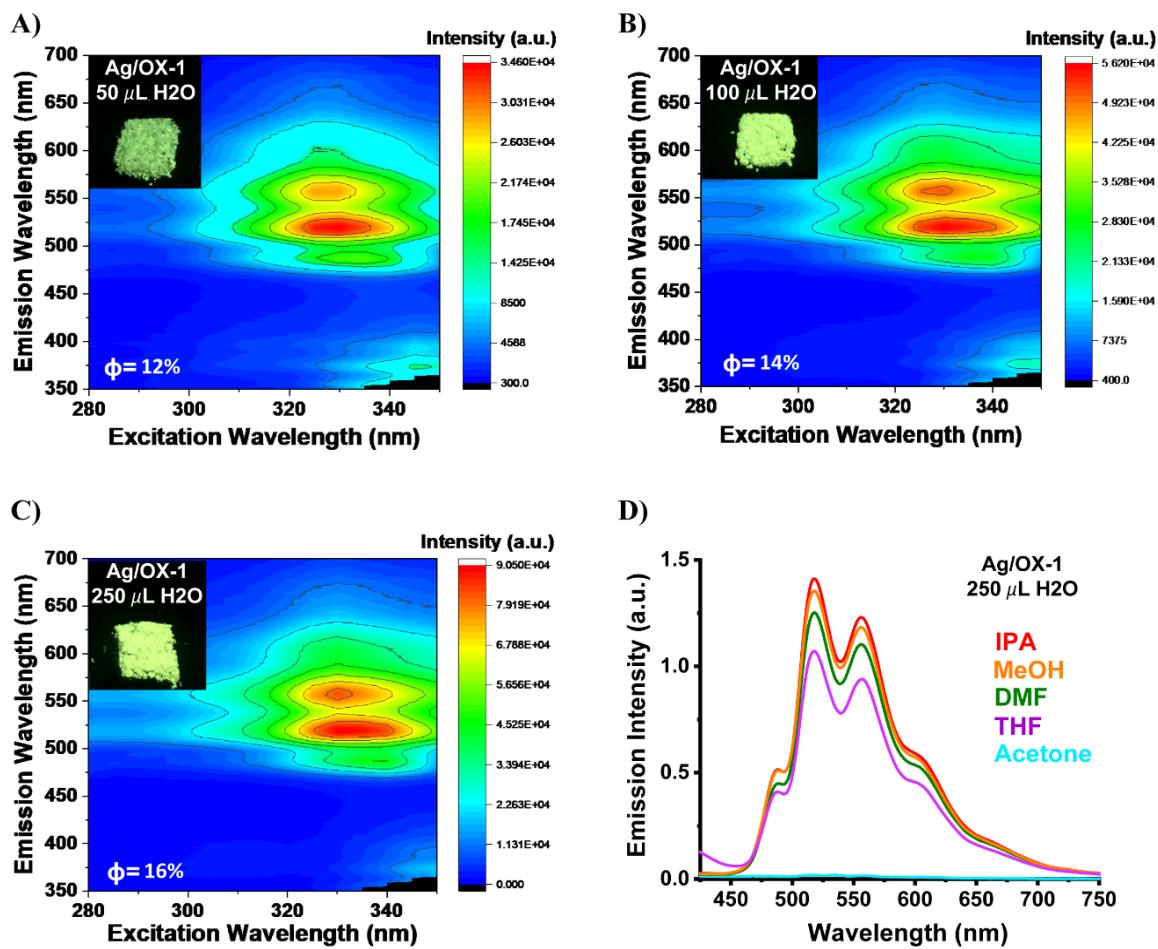

**Figure S10.** Excitation-emission map of **A)** Ag/OX-1 ground at ambient conditions and **B)** Ag/OX-1 ground after the addition of 1 drop of water. The inset displays the emission quantum yield of each sample upon excitation at 330 nm. **C)** Emission spectra of Ag/OX-1 ground with 1 drop of water in different solvents and **D)** is a graphical representation of its intensity maximum in each solvent. The samples were excited at 330 nm.

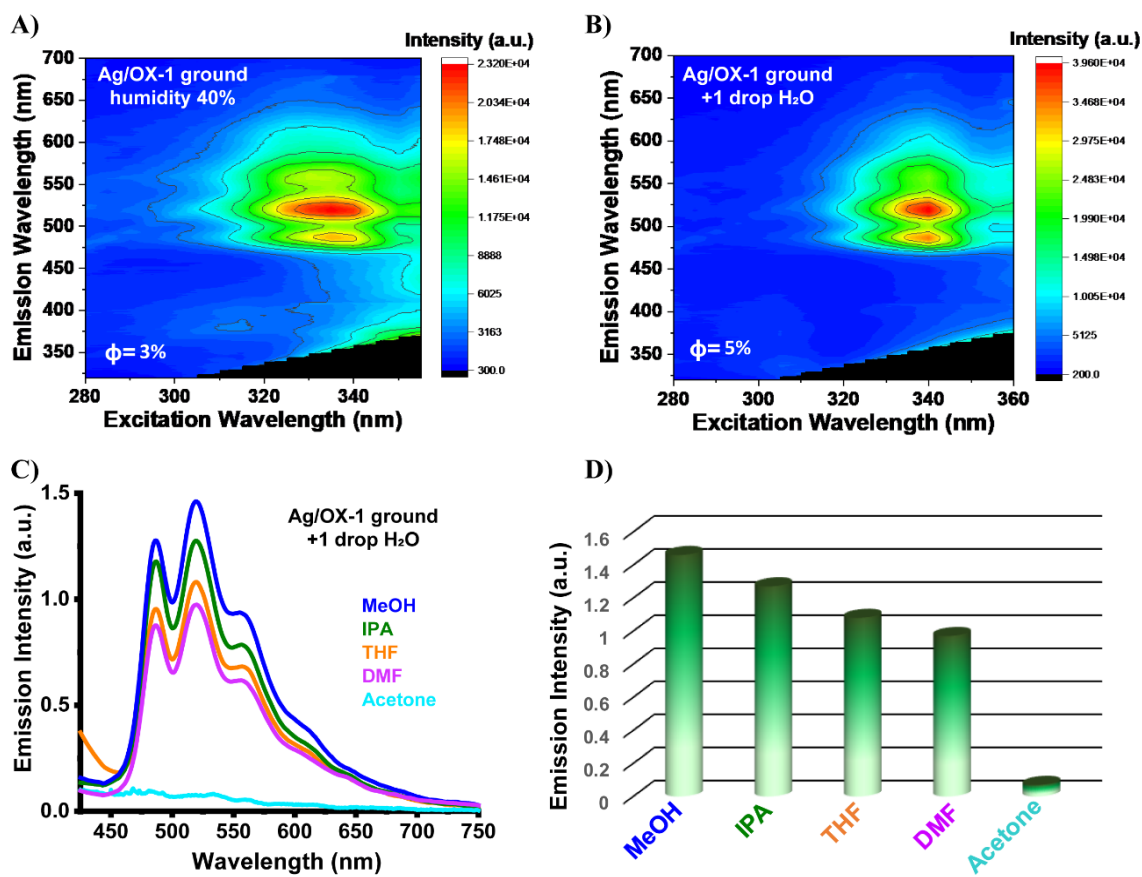

**Figure S11.** PXRD patterns of the OX-1/PU membrane before and after being immersed in water for 20 minutes.

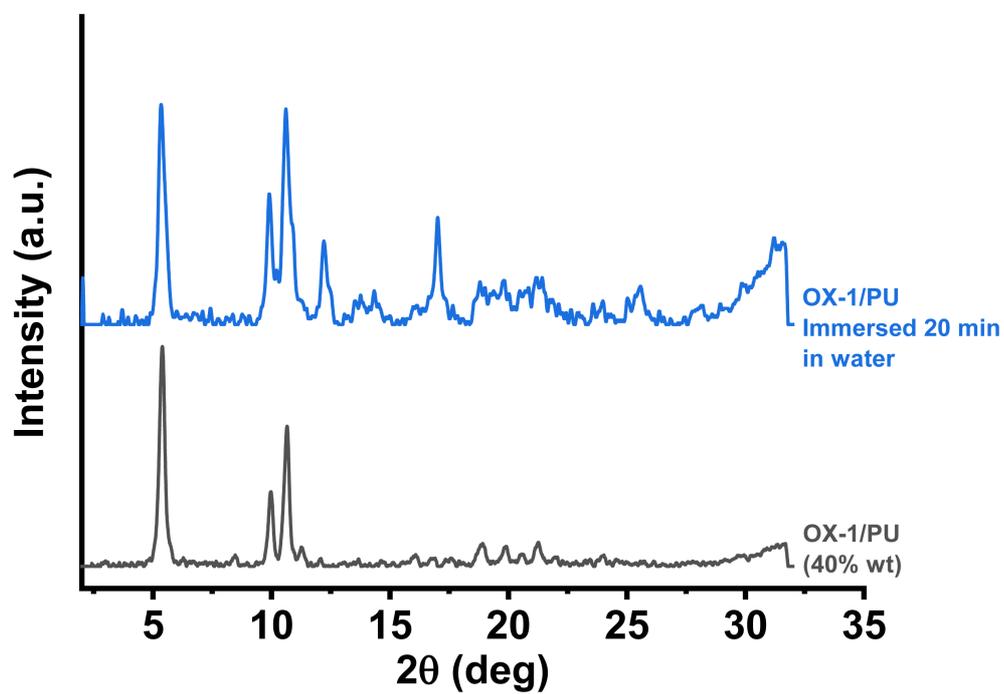